\documentclass[aps,prb,twocolumn,superscriptaddress,showpacs]{revtex4-1}
\usepackage{graphicx}
\usepackage{mathrsfs}
\usepackage{bm}
\usepackage{amsmath}
\usepackage{dcolumn}
\usepackage{epstopdf}
\usepackage{dsfont}
\usepackage{amssymb}
\usepackage{tabularx}
\usepackage{array}
\usepackage{float}
\usepackage{color}
\usepackage{epstopdf}
\usepackage{mathrsfs}
\usepackage[colorlinks, linkcolor=red,anchorcolor=blue,citecolor=green]{hyperref}

\begin{document}

\title{Electric field controlled spin- and valley-polarized edge states in silicene with extrinsic Rashba effect }
%of spin- and valley-polarized edge states in silicene with extrinsic Rashba effect }

\author{Zhiming Yu}
\affiliation{School of Physics, Beijing Institute of Technology, Beijing 100081, China}
\author{Hui Pan}
\affiliation{Department of Physics, Beihang University, Beijing 100191, China}
\author{Yugui Yao}
\email{ygyao@bit.edu.cn}
\affiliation{School of Physics, Beijing Institute of Technology, Beijing 100081, China}

\date{\today}

\begin{abstract}

  In the presence of extrinsic Rashba spin-orbit coupling, we find that silicene can host a new quantum anomalous Hall state with spin- and valley-polarized edge states, which can be effectively controlled by the exchange field and electric field.
  In this new state, the pair of nontrivial  edge states reside in one specific valley and have a strong but opposite spin polarization.
  A distinctive feature of this new state is that both of the spin and valley index of the edge states can be switched by reversing the electric field.
  We also present a microscopic mechanism for the origination of the new state.
  Our findings provide an efficient way to control the topologically protected spin- and valley-polarized edge states, which is crucial for spintronics and valleytronics.

\end{abstract}


\maketitle
\section{Introduction}

Two dimensional topological insulators with honeycomb structure
    have attracted much attention due to various kinds of binary degree of freedom: spin, valley and sublattice\cite{Haldane prl 1988,Zhang prl 2013,Yao  2014,Yao2  2014,Xiao natcomm 2011},
    and the topologically protected metallic edge states\cite{Rev. Kane,Rev. Zhang}.
Nanoelectronic devices designed by topologically protected edge states
    are immune to the  disorder and defects of system.
Furthermore,
    these devices can take advantage of   the intrinsic  spin or valley degrees of freedom of system.
Under some specific situations,
    the  edge states  of honeycomb  materials are spin- or valley-polarized \cite{Qiao2011prl}.
However, if  these materials are planar such as graphene,
    the binary degree of freedom of the edge states are difficult to switch and thus cannot be applied in nanodevices.
On the other hand,
    the electronic properties of low-buckled honeycomb materials
    can be tuned by electric field naturally.

Silicene, which  is a  low-buckled honeycomb material\cite{CC1,CC2},
    exhibits various physical properties
    such as
    quantum spin Hall (QSH) effect\cite{CC1,CC2,kane1,kane2},
    quantum anomalous Hall  (QAH) effect\cite{Ezawa QAH,Ezawa QAH2,Pan},
    strong circular dichroism\cite{Ezawa opt}
    and superconductivity\cite{SupCon,SupCon2}.
It also undergoes a topological phase transition
    by irradiating circular polarized light in the presence of electric field\cite{Ezawa PiQAH,Pi2015}.
Silicene has been synthesized on different substrates,
    including the surface of $Ag(111)$,  $ZrB_{2}$, $In(111)$ \textit{etc}\cite{experiment,experiment2,experiment3,experiment4}.
Moreover,  silicene field-effect transistor has been fabricated
    at room temperature recently\cite{roomT2015}.

Generally, the extrinsic Rashba spin-orbital coupling (SOC)
    of pristine silicene is zero because of the structure inversion symmetry\cite{CC2}.
And the extrinsic Rashba SOC induced by external electric field is  negligible\cite{CC2,Ezawa QAH}.
But strong extrinsic Rashba SOC may arise due to metal-atom adsorption or substrate,
    as they  break the structure inversion symmetry of system dramatically, as discussed in graphene\cite{Dedkov2008,Fe,Qiaoprb2010,Qiaoprb20102}.
Recently, the first-principles calculations have shown  that the strong extrinsic SOC effect in
    silicene with adsorption of different transition metal atoms
    attributes $\sim 7-44\ meV$ to the band gap\cite{ad si},
    which is much larger than the gap ($1.55\ meV$) of pristine silicene\cite{CC1}.
The ferromagnetic substrate and transition metal adatoms also can induce sizable exchange field\cite{Liu 2013,Kaloni 2014}.

In this work,
    we present that silicene with  extrinsic Rashba SOC
    can host a new topological state in the space  of exchange field and  electric field:
    spin- and valley-polarized QAH insulator (SV-QAHI),
    possessing a nonzero Chern number  $\mathcal{C}=\pm 1$
    and     a pair of nontrivial  edge states localized at  a specific valley with a strong but opposite spin polarization.
This new state is different from the
    valley-polarized QAH (VQAH)  insulator reported by Pan et. al.\cite{Pan},
    which is induced by  the competition between intrinsic and extrinsic Rashba SOC.
SV-QAHI  can emerge without  intrinsic Rashba SOC.
A distinctive feature of SV-QAHI is that
    the spin and valley direction of the gapless edge states can be easily  switched
    by reversing the  electric field,
    resulting in  an  electric filed control of spin and valley index.
These switching properties of spin and valley index are useful for spintronics and valleytronics.

The present paper is composed as follows. In Sec. II, we introduce the tight-binding Hamiltonian of silicene with  extrinsic Rashba SOC, exchange field and electric field. Section III presents the phase diagram of silicene in the space of exchange field and electric field. We also discuss the exotic properties and the origin of the nontrivial edge states of SV-QAHI in this section. Finally, we give our conclusion and summarize our results in Sec. IV.

\section{Hamiltonian}

The tight-binding (TB) Hamiltonian of silicene with  extrinsic Rashba SOC, exchange field and electric field  reads\cite{CC2}:
\begin{eqnarray}
H & = & -t\sum_{<i,j>\alpha}c_{i\alpha}^{\dagger}c_{j\alpha}+i\ \lambda\sum_{<i,j>\alpha\beta}c_{i\alpha}^{\dagger}\left(\boldsymbol{\sigma\times d_{ij}}\right)_{\alpha\beta}^{z}c_{j\beta}\nonumber \\
 &  & +M\sum_{i\alpha\beta}c_{i\alpha}^{\dagger}\sigma_{\alpha\beta}^{z}c_{i\beta}+\Delta\sum_{i\alpha}\mu_{i}c_{i\alpha}^{\dagger}c_{i\alpha}\label{eq:Ham1}
\end{eqnarray}
where $c_{i\alpha}^{\dagger}$($c_{i\alpha}$) is the electronic creation (annihilation) operator with spin $\alpha$ at site $i$.
    $\boldsymbol{\sigma}$ is the Pauli matrix,
    $\boldsymbol{d}_{ij}$ represents a unit vector connecting sites $i$ and $j$,
    and $<i,j>$ stands for nearest-neighbor sites.
The first term is the usual nearest-neighbor hopping with $t=1.08\ eV$\cite{CC2}.
The second and third  terms represent the extrinsic Rashba SOC ($\lambda$) and exchange field ($M$)  respectively,
     both can induced by ferromagnetic substrate or adsorption of transition metal atoms on silicene\cite{ad si}.
$\Delta$ is the staggered sublattice potential arising from
    an electric field perpendicular to silicene sheet.
Here, we neglect the influence of electric field on extrinsic Rashba SOC.
We also neglect the intrinsic SOC  and intrinsic Rashba SOC,
    as the adatoms or substrate induced extrinsic Rashba SOC  may be much stronger  than the intrinsic SOCs\cite{Dedkov2008,Qiaoprb2010,ad si},
    in  which situations  we are interested.

\section{Phase diagram and spin- and valley-polarized edge states}

\begin{figure}[t]
\includegraphics[scale=0.75]{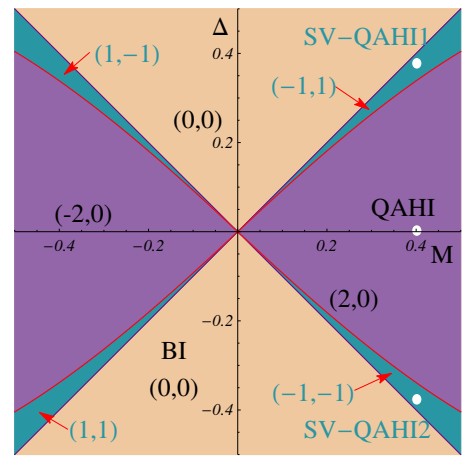}

\protect\caption{
(Color online) Phase Diagram of silicene in the $(M,\Delta)$ space with extrinsic Rshaba SOC $\lambda=0.2\ t$, calculated from the TB Hamiltonian (\ref{eq:Ham1}).
The solid lines separate the phases,
    which  are indexed by the Chern number and  the valley index of the edge state $({\cal{C}},\ V_{edge})$.
    We define $V_{edge}\equiv1/-1$   when the gapless edge states are located around   $K/K^{\prime}$ valley,
    and  $V_{edge}\equiv0$ if  there is no gapless edge state     or  the gapless edge states are
    not located around  one valley.
$M$ and $\Delta$ are in unit  of $t$. \label{fig:Phase-Diagram}}
\end{figure}

\begin{figure*}[htbp]
\includegraphics[scale=0.75]{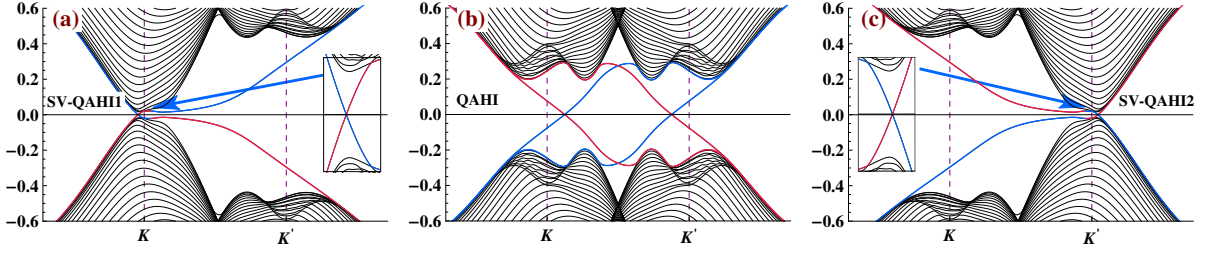}

\protect\caption{
(Color online) Band structures of the zigzag-terminated nanoribbon at marked points in Fig. \ref{fig:Phase-Diagram}, calculated from the TB Hamiltonian (\ref{eq:Ham1}).
The topological  index $({\cal{C}},\ V_{edge})$ are
    (a) SV-QAHI1:  $(1,1)$ ($\Delta=0.38\ t$ $\&$ $M=0.4\ t$);
    (b) QAHI: $(2,0)$ ($\Delta=0\ t$ $\&$ $M=0.4\ t$);
    (c) SV-QAHI2: $(1,-1)$ ($\Delta=-0.38\ t$ $\&$ $M=0.4\ t$).
Blue/Red curves   represents edge states propagating  along  the top/bottom  edge of nanoribbon.
Energy is in unit of $t$ and momentum is in unit of $1/a$.\label{fig:Band-ribbon}}
\end{figure*}

In Fig. \ref{fig:Phase-Diagram},
    we presented the phase diagram of silicene in the $(M,\Delta)$ space.
All the phases are characterized by the Chern number ($\cal{C}$) and the valley degree of freedom of the nontrivial edge states ($V_{edge}$).
The Chern number ($\cal{C}$) determines the topological properties of phase.
And $V_{edge}$ is introduced to present the valley-polarized properties of the nontrivial edge states.
We define $V_{edge}\equiv 1/-1$ when the gapless edge states are localized around $K/K^{\prime}$ valley,
    and  $V_{edge}\equiv 0$ if  there is no gapless edge states
    or  the gapless edge states are not localized around  one specific valley.
When  exchange field  dominates  over the electric field,
    silicene is a quantum anomalous Hall insulator (QAHI) with $\left({\cal C},V_{edge}\right)=\left(2,\ 0\right)$
    for $M>0$.
 In Fig. \ref{fig:Band-ribbon},
    we present the band structures of the zigzag-terminated nanoribbon  at marked points in Fig. \ref{fig:Phase-Diagram}.
The nanoribbon band structure of QAHI with $\left({\cal C},V_{edge}\right)=\left(2,\ 0\right)$ is presented in Fig. \ref{fig:Band-ribbon}b.
Silicene  undergoes a topological phase transition from QAHI to  SV-QAHI
    when the strength of external electric field increases
    and crosses a critical value.
The  index of SV-QAHI is $\left(-1,1\right)$
    for $M>0$ and $\Delta>0$, as shown  in Fig. \ref{fig:Band-ribbon}a.
Keep increasing the electric field,
    silicene would undergo another topological phase transition to
    a band insulator (BI) with  index $\left(0,\ 0\right)$.

\begin{figure}[ht]
\includegraphics[scale=0.75]{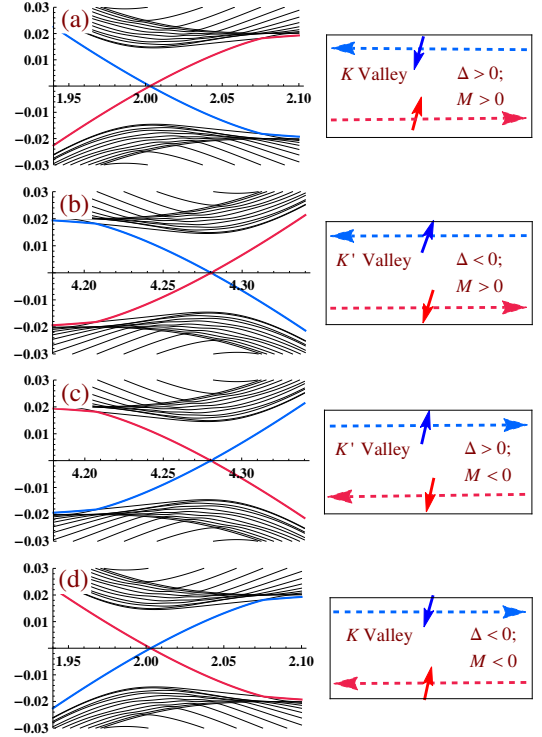}
\protect\caption{(Color online) Enlarged edge states of zigzag-terminated nanoribbons of SV-QAHI in the four  quadrant of  $(M,\Delta)$ space.
Strong  spin polarization edge currents propagate along the nanoribbons edges.
Blue/Red curves   represents edge states propagating  along  the top/bottom  edge of nanoribbon.
By reversing  the external electric field ($\Delta$) or  exchange field ($M$), the polarization direction of spin and valley  of the edge states can be switched. \label{fig:filter}}
\end{figure}

The most exciting finding is that the edge  states of zigzag-terminated  SV-QAHI are approximately mono-component,
    and the spin and valley direction  of them can be switched by electric field.
For clarity,
    we define  the uppermost/downmost atoms of the zigzag-terminated  nanoribbon belong to sublattice A/B.
Let us start from SV-QAHI  with  index $\left(-1,1\right)$ ($\Delta>0$ $\&$ $M>0$),
    of which  the edge currents  at the top- and bottom-edge of nanoribbon are  mainly composed by $|A,K,\downarrow>$ and $|B,K,\uparrow>$
     respectively, as shown in Fig. \ref{fig:filter}a.
If we reverse the electric field  and keep exchange field unchanged ($\Delta<0$ $\&$ $M>0$),
    silicene becomes a $\left(-1,-1\right)$ SV-QAHI (Fig. \ref{fig:filter}b),
    and the binary indices of top- and bottom-edge states  become $|A,K^{\prime},\uparrow>$ and $|B,K^{\prime},\downarrow>$,
    directly demonstrating  that  the spin and valley index of the edge states have been switched.
And if we reverse the exchange field instead of electric field ($\Delta>0$ $\&$ $M<0$),
    silicene turns into $(1,-1)$ SV-QAHI (Fig. \ref{fig:filter}c),
    and the binary  indices of top- and bottom-edge states
    become $|A,K^{\prime},\uparrow>$ and $|B,K^{\prime},\downarrow>$ respectively
    which are same as that of $\left(-1,-1\right)$ SV-QAHI (Fig. \ref{fig:filter}b).
But the direction of top- and bottom-edge current   between  $(1,-1)$ and  $\left(-1,-1\right)$  SV-QAHI are opposite,
    which can be found by the Chern numbers and
    the slopes of the edge states of  these two insulators.
Similarly,
    the binary indices of the edge currents of $\left(1,1\right)$
    SV-QAHI (Fig. \ref{fig:filter}d) are the same as that of $\left(-1,1\right)$ SV-QAHI (Fig. \ref{fig:filter}a),
    while the direction of edge currents of them are opposite too.
In short,
        we can switch the polarization direction of spin and valley  of the edge states by reversing electric field.
And the polarization direction  of  spin, valley and the  propagation  of  edge current are all switched by reversing exchange field.

In order to understand the underlying physics of  SV-QAHI,
    we first explore the phase diagram of silicene analytically  by a low-energy effective Hamiltonian\cite{Naka},
    then discuss the origination  of the approximately mono-component edge states,
    and the electric field control of spin and valley direction of the edge states.
First,
we expand the TB Hamiltonian (\ref{eq:Ham1}) around  valley points ($K$ and $K^{\prime}$):
\begin{eqnarray}
H^{\tau} & = & \left(\begin{array}{cccc}
\Delta+M & h_{+} & 0 & -v_{F}k_{+}^{\tau}\\
h_{+}^{*} & -\Delta-M & -v_{F}k_{-}^{\tau} & 0\\
0 & -v_{F}k_{+}^{\tau} & \Delta-M & h_{-}\\
-v_{F}k_{-}^{\tau} & 0 & h_{-}^{*} & -\Delta+M
\end{array}\right)\label{eq:hamval}
\end{eqnarray}
with the basis $\left(A\uparrow,B\downarrow,A\downarrow,B\uparrow\right)^{T}$, and
\begin{eqnarray}
h_{\pm} & = & i\frac{3}{2}\lambda\left(1\pm\tau\right)-\frac{\sqrt{3}}{4}\lambda\left(\tau\mp1\right)\left(ik_{x}+\tau k_{y}\right)a \label{eq:hpm}
\end{eqnarray}
    where $\tau=\pm1$ represents valley degrees of freedom $K$($K^{\prime}$),
    and $k_{\pm}^{\tau}=\tau k_{x}\pm ik_{y}$.
$v_{F}=\frac{\sqrt{3}}{2}at=5.5\times10^{5}\ m/s$
    is Fermi velocity with the lattice constant $a=3.86\ {\AA}$\cite{CC2}.
Since the extrinsic Rashba SOC is not negligible in the present model,
    we keep $\lambda k_{x(y)}$  in  the second term of Eq. (\ref{eq:hpm}),
    and it turns out that $\lambda k_{x(y)}$ is crucial to obtain the SV-QAHI,
    and  leads to the trigonal warping effects around the valleys.
As shown in Eq. (\ref{eq:hpm}),
    the extrinsic Rashba SOC term is strongly valley dependent,
    and couples $|A\uparrow>$($|A\downarrow>$) and $|B\downarrow>$($|B\uparrow>$)
    as to  make the spin and sublattice degrees of freedom are tightly connected to each other.
Therefore, the operations on sublattice potential by electric field would affect the spin,
    which gives rise to a possibility of full electric-field control of spin of the system.

The energy bands  of silicene at $K$ or $K^{\prime}$ points are $\pm (\Delta+M)$ and $\pm\sqrt{(M-\Delta)^{2}+9\lambda^{2}}$ or $\pm(\Delta-M)$ and $\pm\sqrt{(M+\Delta)^{2}+9\lambda^{2}}$, respectively.
When $\Delta$ approaches $M$,
        the low-energy bands ($\pm (|\Delta|-|M|)$) are  well separated from the other high-energy bands ($\pm\sqrt{(|M|-|\Delta|)^{2}+9\lambda^{2}}$) owing to nonzero $\lambda$.
Thus, in the presence of extrinsic Rashba SOC and $|\Delta| \simeq |M|$,
        a   two-band low energy effective Hamiltonian  can be established\cite{CC2}:
\begin{eqnarray}
H_{eff}(\boldsymbol{k}) & = & \boldsymbol{d}(\boldsymbol{k})\cdot\boldsymbol{\sigma}\label{eq:eff}
\end{eqnarray}
with
\begin{eqnarray*}
d_{x} & = & -\lambda k_{y}a\left(\frac{\sqrt{3}}{2}+3\eta\Gamma k_{x}/a\right)\\
d_{y} & = & \frac{\sqrt{3}\lambda}{2}\left(\eta k_{x}a-\sqrt{3}\Gamma(k_{x}^{2}-k_{y}^{2})\right)\\
d_{z} & = & \Delta-\eta M\left(1-\Gamma\boldsymbol{k}^{2}\right)
\end{eqnarray*}
where $\eta=\text{Sgn}(M\Delta)$ and $\Gamma=2v_{F}^{2}/\left(4|\Delta M|+9\lambda^{2}\right)$.
When $\eta=1/-1$, the low-energy
    states are localized around  $K$/$K^{\prime}$ valley,
    as shown  in Fig. \ref{fig:filter}.
The basis of $H_{eff}(\boldsymbol{k})$ is
    $\left(A\downarrow,B\uparrow\right)^{T}$ when $\eta=1$
    and $\left(A\uparrow,B\downarrow\right)^{T}$ when $\eta=-1$.
%The  second order of $k_{x(y)}$ in $\boldsymbol{d(k)}$ leads to the trigonal warping effects around the valley.
The effective Hamiltonian also satisfies the following symmetry operations,
\begin{eqnarray}
H_{eff}(-\Delta,M,\boldsymbol{k})& = &\sigma_{y}H_{eff}(\Delta,M,-\boldsymbol{k})\sigma_{y} \label{eq:symm1} \\
 H_{eff}(-\Delta,-M,k_{y}) & = & \sigma_{y}H_{eff}(\Delta,M,-k_{y})\sigma_{y} \label{eq:symm2}
\end{eqnarray}
Based on these symmetry operations,
    we can assume $\Delta>0$ and $M>0$ in the following discussion
    without loss of generality.
The energy spectrum can be directly  derived as $\varepsilon_{\boldsymbol{k}}=\pm|\boldsymbol{d}|$.
The zero energy $\varepsilon_{k}=0$ only occurs when all the elements of $H_{eff}(\boldsymbol{k})$ vanish.
We obtain two  phase boundaries:
    $\Delta_{+}(M)=M$
    with single Dirac point located at $K$ point $(\boldsymbol{k}_{+}=0)$,
    and
\begin{eqnarray}
\Delta_{-}(M) & = & M\frac{6v_{F}^{2}-9\lambda^{2}a^{2}}{6v_{F}^{2}+4M^{2}a^{2}}\label{eq:Delta}
\end{eqnarray}
with  three Dirac points located at
\begin{eqnarray}
\boldsymbol{k}_{-} & = & k_{-}e^{in\pi/3}=\frac{a\left(4|\Delta_{-} M|+9\lambda^{2}\right)}{2\sqrt{3}v_{F}^{2}}e^{in\pi/3} \label{eq:k-}
\end{eqnarray}
with $n=0,\ \pm 1$ corresponding to the trigonal warping terms.
The SV-QAHI exists over the electric field region $\Delta_{-}<\Delta<\Delta_{+}$.
Note that  $H_{eff}(\boldsymbol{k})$ is only well defined around the valley centers,
        hence,
        $\Delta_{-}(M)$   (Eq. (\ref{eq:Delta})) would be invalid
        when $\lambda$ or $M$  is large, where $k_{-}$ (Eq. (\ref{eq:k-})) is large too.
However, SV-QAHI does emerge when   $\Delta$ is smaller than $M$ and larger than a critical value,
        which can be found by  the calculations of TB Hamiltonian (\ref{eq:Ham1}).

\begin{figure}[t]
\includegraphics[scale=0.43]{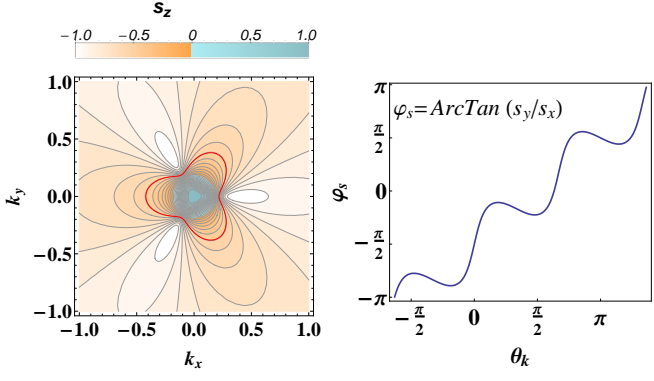}

\protect\caption{(Color online)
Skyrmion spin  texture needs $s_{z}$ ranges from $1$ to $-1$ and the  azimuth angle of spin $\varphi_{s}=ArcTan\left(s_{y}(\boldsymbol{k})/s_{x}(\boldsymbol{k})\right)$ changes from  $0$ to $2\pi$ with any  $s_{z}$.
Left:
The contour map of $s_{z}$ around $K$ valley of SV-QAHI for $M>0$ and $\Delta>0$, calculated from the effective Hamiltonian $H_{eff}(\boldsymbol{k})$ (\ref{eq:eff}),
    which shows the reversal of spin  and the trigonal warping effect around valley.
Right: Along the isoline of $s_{z}(\boldsymbol{k})$,
the azimuth angle of spin  $\varphi_{s}$  changes from $0$ to $2\pi$.
Here we present the behavior of $\varphi_{s}$
as a function of $\theta_{k}=ArcTan\left(k_{y}/k_{x}\right)$ with
$s_{z}(\boldsymbol{k})=-0.6$.
The isoline of $s_{z}(\boldsymbol{k})=-0.6$
is marked by red curve in left figure.
With other  $s_{z}(\boldsymbol{k})$, $\varphi_{s}$ also presents similar behavior (not shown).
Hence, these two figures directly indicate that $H_{eff}(\boldsymbol{k})$ generates a Skyrmion spin  texture. \label{fig:sky}}
\end{figure}

Now we study the physical properties of the  gapless edge states of  SV-QAHI by the  effective
    Hamiltonian $H_{eff}(\boldsymbol{k})$.
Generally, the topological properties of system can also be understood by Skyrmion spin (sublattice) texture\cite{Qiao 2012prb,Pan 2015prb}.
In Fig. \ref{fig:sky}, we present the contour map of  z-component spin (sublattice) ($s_{z}(\boldsymbol{k})$) of
    SV-QAHI in the vicinity of  Fermi level.
$s_{z}(\boldsymbol{k})=<v_{k}|\sigma_{z}|v_{k}>$
    with $|v_{k}>=\frac{1}{N_{k}}\left(d_{z}-|\boldsymbol{d}|,\ d_{x}+id_{y}\right)^{T}$
    being the eigenfunction of valence band of $H_{eff}(\boldsymbol{k})$
    and $N_{k}$ the normalization constant.
The expression of $s_{z}(\boldsymbol{k})$
    is too long and not necessary to present here.
The important things are that in the SV-QAHI phase,
    $s_{z}(\boldsymbol{k}=0)=1$
    and $s_{z}(\boldsymbol{k}^{\prime})=-1$
    with $\boldsymbol{k}^{\prime}=ae^{in\pi/3}/\left(\sqrt{3}\Gamma(\Delta)\right)$
    and $n=0,\pm 1$.
Along the isoline of $s_{z}(\boldsymbol{k})$,
    the azimuth angle of spin $\varphi=ArcTan\left(s_{y}(\boldsymbol{k})/s_{x}(\boldsymbol{k})\right)$
    changes from $0$ to $2\pi$ continuously, which is shown in Fig. \ref{fig:sky} too.
Therefore a Skyrmion spin (sublattice) texture is generated   in the  SV-QAHI phase.

\begin{figure}[h]
\includegraphics[scale=0.8]{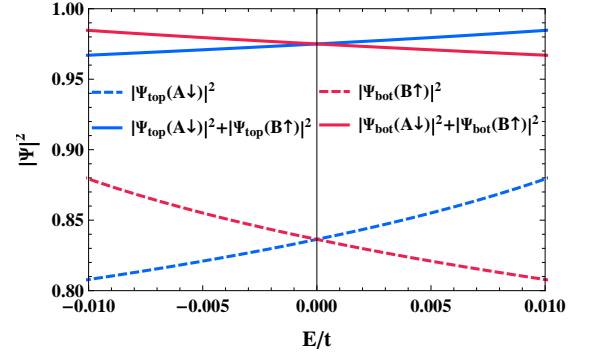}

\protect\caption{
(Color online) The probability of $|A\downarrow>$ and $|B\uparrow>$ in   the edge states of  zigzag-terminated  SV-QAHI nanoribbon ($M>0$ and $\Delta>0$),
        obtained from the TB Hamiltonian (\ref{eq:Ham1}).
 The solid lines intuitively show the edge states are almost composed by the basis of $H_{eff}(\boldsymbol{k})$:
        $\left(A\downarrow,\ B\uparrow\right)^{T}$.
The dash lines show that
    the edge current propagating   along top and bottom edge have strong  spin polarization:
    mainly composed by $|A\downarrow>$ and $|B\uparrow>$ respectively.
\label{fig:edge state }}
\end{figure}

Usually, Skyrmion and nontrivial edge states are two sides of the topological properties of system.
Thus,  the nontrivial  edge states of nanoribbon can be expected to be mainly composed by $|A\downarrow>$  and  $|B\uparrow>$,
        which are the basis of the low-energy effective Hamiltonian $H_{eff}(\boldsymbol{k})$.
Fig. \ref{fig:edge state } presents the probability of
        $|A\downarrow>$ and $|B\uparrow>$ in the  gapless edge states within the bulk energy gap.
It shows that
        the total  probability of $|A\downarrow>$ and $|B\uparrow>$ in the  edge states is more than $96\%$ (solid curves in Fig. \ref{fig:edge state }).
And the top/bottom edge current is   mainly  composed by $|A\downarrow>$/$|B\uparrow>$ (dashed curves in Fig. \ref{fig:edge state }).
 From the effective Hamiltonian $H_{eff}(\boldsymbol{k})$,
    we know that if we reverse the electric field,
    the position of low-energy states changes from $K$ valley to $K^{\prime}$ valley
    and the  basis of $H_{eff}(\boldsymbol{k})$
    changes from $\left(A\downarrow,B\uparrow\right)^{T}$ to   $\left(A\uparrow,B\downarrow\right)^{T}$.
Consequently,
       by reversing electric field,
        the  component of the edge current  propagating along  the top/bottom-edge changes from $|A\downarrow>$/$|B\uparrow>$ to  $|A\uparrow>$/$|B\downarrow>$,
        resulting in the  electric field switching of spin and valley direction of the edge states.

At last, we make some discussions  on the realization  of  SV-QAHI.
 From the effective Hamiltonian $H_{eff}(\boldsymbol{k})$ ({\ref{eq:eff}}),
        we note that the SV-QAHI can exist over a wide electric field region
        when $M$ or $\lambda$ is large.
Hence, the SV-QAHI may appear and be observed in silicene with suitable adsorption of transition metal atoms and ferromagnetic substrate.
Moreover, since germanene and stanene share most of the electronic properties of silicene,
        we expect this SV-QAHI also can emerge in these two materials\cite{Kaloni2014}.

%From  Fig. \ref{fig:Phase-Diagram} and the effective Hamiltonian $H_{eff}(\boldsymbol{k})$ ({\ref{eq:eff}}),
%        we note that the SV-QAHI can be   obtained by tuning electric field
%        even if  the strength of  exchange field and extrinsic Rashba SOC are  small.
%The SV-QAHI would exist over a large electric field region
%    when $M$ or $\lambda$ is large,
%    which may be the situation of silicene  with adsorption of transition metal atoms\cite{ad si}.
%Hence, in these adsorbed materials, the SV-QAHI may appear and be observable.

\section{Conclusions}
In summary,
        we present that a new  quantum anomalous Hall state
        can be obtained in silicene with  extrinsic Rashba SOC.
The gapless edge states of a zigzag-terminated nanoribbon of this new state
        have strong but opposite spin polarization and are  located around  one specific valley.
The emergence  of the novel spin- and valley-polarized edge states are explained in detail
        by two band  low-energy effective Hamiltonian model.
It is found that the extrinsic Rashba SOC is crucial for realizing such new states.
Particularly,
        the direction of spin and valley of  the edge states can be  easily switched by reversing  the   electric field or  exchange field.
This electric field control of spin and valley provides an opportunity to design
        newly  topological-state based spintronic and valleytronic devices.

%\section{acknowledgments}
\begin{acknowledgments}
We are thankful to Shengyuan A. Yang  for  revising this manuscript.
This work was supported by the MOST Project of China (Grants Nos. 2014CB920903, and 2011CBA00100) , the NSFC (Grants Nos.
11174022, 11174337, 11225418, 61227902, and 11574029), and the Specialized Research Fund for the Doctoral Program of Higher Education of China (Grant No. 20121101110046).

\end{acknowledgments}


\end{document}